\documentstyle[12pt,subeqnarray, cite]{article}

\begin{document}
\newcommand{\dis}{\displaystyle}
\newcommand{\id}{ 1 \hspace{-2.85pt} {\rm I} \hspace{2.5mm}}
\makeatletter
\renewcommand{\theenumi}{\roman{enumi}}
\renewcommand{\labelenumi}{(\theenumi)}
\renewcommand{\p@enumii}{\theenumi--}
\makeatother
{\pagestyle{empty}
\rightline{} 
\rightline{} 
\rightline{} 
\vskip 1cm
\centerline{\large \bf Two-State Spectral-Free   }
\centerline{\large \bf Solutions of }
\centerline{\large \bf Frenkel-Moore Simplex Equation.}

\vskip 2cm
\centerline{ L. C. Kwek {\footnote{E-mail address:
SCIP3057@LEONIS.NUS.SG }},  C. H. Oh {\footnote{E-mail address:
PHYOHCH@NUS.SG }} , K. Singh  and K.Y. Wee.
}
\centerline{{\it Department of Physics, Faculty of Science, } }
\centerline{{\it National University of Singapore,Lower Kent Ridge,} }
\centerline{{\it Singapore 0511, Republic of Singapore. } }
\vskip 0.1in

\vskip 1cm
\centerline{\bf Abstract} \vspace{10mm}

\noindent{Whilst many solutions have been found for the Quantum
Yang-Baxter Equation (QYBE), there are fewer known solutions available
for its higher dimensional generalizations: Zamolodchikov's tetrahedron
equation (ZTE) and Frenkel and Moore's simplex equation (FME).  In this
paper, we present families of solutions to FME which may help us to
understand more about higher dimensional generalization of QYBE.}

\newpage
}

\section{Introduction}

\hspace*{0.2in} The Quantum Yang-Baxter Equation (QYBE) plays a pivotal
role in the study of two-dimensional integrable models, quantum groups,
conformal field theory and the study of link polynomials in knot
theory.  A systematic study of the solutions of the Quantum Yang-Baxter
Equation shows that there are an infinite number of two-dimensional
exactly solvable models in classical statistical mechanics
\cite{wadati1, sogo}.

Using a computer algebra method, Hietarinta \cite{hieta1, hieta2} has
obtained the complete classification of all two-state solutions of the
QYBE.  He has also extended his work in search of three-state solutions of
the constant YBE \cite{hieta4} by studying all upper triangular ansatze.

Higher dimensional generalization of QYBE is possible. By considering
the scattering amplitudes of straight strings in $2 + 1$ dimensions,
Zamolodchikov \cite{zamo1, zamo2} derives a three-dimensional equivalent of
QYBE, commonly called the tetrahedron equation (ZTE):
\begin{eqnarray}
& & R_{123}(\theta_{1}, \theta_{2}, \theta_{3})R_{145}(\theta_{1},
\theta_{4}, \theta_{5})R_{246}(\theta_{2},\theta_{4},\theta_{6})
R_{356}(\theta_{3},\theta_{5},\theta_{6})  \nonumber \\
& = &
R_{356}(\theta_{3},\theta_{5},\theta_{6})
R_{246}(\theta_{2},\theta_{4},\theta_{6})R_{145}(\theta_{1},
\theta_{4}, \theta_{5})R_{123}(\theta_{1}, \theta_{2}, \theta_{3}),
\end{eqnarray}
where $R_{123} = R \otimes \id$ etc and $R \in \ {\rm End}(V
\otimes V \otimes V) $ for some vector space $V$. In the same paper, he
also  ingeniously provided a non-trivial spectral dependent two-state
solution. Baxter \cite{baxter} subsequently proved that the solution
conjectured in Zamolodchikov's paper satisfies the tetrahedron
equation. The tetrahedron equation is by no means simple.  Even in the
two-state case, there are $2^{14}$ consistency equations with $2^{6}$
variables. An N-state generalization of the tetrahedron solution has
also been found using a free-fermion model on the three dimensional
lattice \cite{baz1,baz2}.

The tetrahedron equation is not the only higher dimensional
generalization. By investigating the symmetry inherent in QYBE, Frenkel
and Moore \cite{frenkel} suggested another possible generalization (FME):
\begin{equation}
R_{123} R_{124} R_{134} R_{234} =
R_{234} R_{134} R_{124} R_{123},
\end{equation}
where $R_{123} = R \otimes \id$ etc and $R \in \ {\rm End}(V
\otimes V \otimes V)$ for some vector space $V$.
While much work 
\cite{baz1, baz2, kash, korep1, korep2, manga1, manga2}
has been done to relate Zamolodchikov's tetrahedron equation to
three-dimenional lattice models, less effort \cite{oh, zheng} has so
far was directed at the Frenkel and Moore generalization.

In section 2, we describe some known symmetries associated with
FME, and review other works done on FME.  We also show that,
unlike ZTE, a cross-diagonal ansatz always satisfies 2-state FME. In
section 3, we briefly describe the technique in our work and then
present our results in section 4.

\section{Frenkel-Moore Simplex Equation}

\subsection{General symmetries}\label{gensym}
\hspace*{0.2in} There are some significant differences between
Zamolodchikov's tetrahedron equation and Frenkel and Moore's simplex
equation.  Essentially, the underlying vector spaces on which the
operator $R$ acts differ. Further, the operator $R$ in Zamolodchikov's
tetrahedron equation  seems to be local whilst the operator $R$ in
Frenkel and Moore's version possesses a global labelling scheme
\cite{frenkel, hieta3}.

By choosing an appropriate basis for $V$, Frenkel and Moore's equation
becomes:
\begin{equation}
R_{abc}^{i_{1}i_{2}i_{3}} R_{i_{1}i_{2}d}^{i_{4}i_{5}i_{6}}
R_{i_{4}i_{3}i_{6}}^{ei_{7}i_{8}} R_{i_{5}i_{7}i_{8}}^{fgh}
= R_{bcd}^{i_{1}i_{2}i_{3}} R_{ai_{2}i_{3}}^{i_{4}i_{5}i_{6}}
R_{i_{4}i_{1}i_{6}}^{i_{7}i_{8}h} R_{i_{7}i_{8}i_{5}}^{efg}
\end{equation}
where repeated indices are summed over.

The equation does not possess a spectral parameter but it is invariant
under similarity transformation \cite{hieta2, oh} similar to QYBE:

\begin{equation}
R \rightarrow \kappa (Q \otimes Q \otimes Q) R (Q^{-1} \otimes Q^{-1}
\otimes Q^{-1})
\end{equation}

\noindent for some non-singular matrix $Q \in End(V) $.
It is also invariant under permutation of the
indices, namely,

\begin{subeqnarray}
R_{ijk}^{lmn} &\rightarrow & R_{(i+r) ~ {\rm mod} \ d \ \ (~ j+r) ~ {\rm mod}
\ d
\ \ (k+r) ~
{\rm mod}\ d}^{(l+r) ~ {\rm mod} \ d \ \ (m+r) ~ {\rm mod} \ d \ \
(n+r) ~ {\rm mod} \ d} \\
R_{ijk}^{lmn} &\rightarrow & R_{lmn}^{ijk} \label{sym2}\\
R_{ijk}^{lmn} &\rightarrow & R_{kji}^{nml} \label{sym3}
\end{subeqnarray}

\noindent where $r = 1, 2, 3, \ldots, d - 1 $, and $dim(V) = d$.
This is known as discrete symmetry \cite{hieta2}

The symmetry transformations  \ref{sym2}(b) and \ref{sym3}(c) imply the
symmetry transformation:
\begin{eqnarray}
R_{ijk}^{lmn} &\rightarrow & R_{nml}^{kji}
\end{eqnarray}

\noindent Indeed, if we consider all possible transformations involving
permutation of the indices $\{ i, j, k, l, m, n \}$,
\ref{sym2}(b), \ref{sym3}(c) are the only ones that will allow FME to remain
invariant. In this paper, we only consider the case when $dim (V)
= 2$.

\subsection{Other works on FME}

\hspace*{0.2in} By considering total symmetric ansatz, namely,
R-matrices in which
\begin{equation}
R_{ijk}^{lmn} = R_{jik}^{mln} = R_{ikj}^{lnm} = R_{kji}^{nml} =
R_{jki}^{mnl} = R_{kij}^{nlm},
\end{equation}
the authors of ref \cite{oh} have successfully listed all totally
symmetric solutions of FME.  They have found five independent solutions
after eliminating those solutions which are related to each other under
symmetry transformations.

Zheng and Zhang \cite{zheng} have also constructed some beautiful
solutions of FME.  They considered an ansatz of the form $\dis \left(
\begin{array}{cc}
X & Y \\
0 & Z \\
\end{array}
\right ) $, where $X$ and $Z$ are solutions of QYBE related to the
superalgebra, the Temperly-Lieb algebra and the Birman-Wenzl algebra.

\subsection{Cross-Diagonal Ansatz}\label{cross}

\hspace*{0.2in} Any diagonal ansatz, which is just an R-matrix with
only diagonal entries,  will satisfy a simplex equation, be it ZTE or
FME. However, the statement is not true if we consider cross diagonal
R-matrices.  A cross diagonal R-matrix is one in which the only
non-zero elements are $R_{ijk}^{(i+1) {\rm mod 2} \ \ (j+1) {\rm mod 2}
\ \ (k+1) {\rm mod 2}}$.  Whilst this ansatz does not necessarily
satisfy the tetrahedron equation unless certain conditions hold, it
will always satisfy the two-state FME.

To see this fact, we simply consider  $R_{ijk}^{\mu \nu \sigma} \neq 0
$ provided $i = \bar{\mu}, j = \bar{\nu}, k = \bar{\sigma}$, where $\mu
+ \bar{\mu} = 1 (mod ~ 2)$ , $\bar\mu$ being the complement of $\mu$.
Substituting into FME, we see that the only non-zero terms on the left
and right hand side of the equation are:
\begin{equation}
R_{abc}^{\bar{a}\bar{b}\bar{c}}R_{\bar{a}\bar{b}d}^{ab\bar{c}}
R_{a\bar{c}\bar{d}}^{\bar{a}cd}R_{bcd}^{\bar{b}\bar{c}\bar{d}}.
\end{equation}

\noindent In contrast, if we substitute this form of the R-matrix into
ZTE, the left and right side of the equation do not necessarily cancel,
and we require the terms
\begin{equation}
R_{abc}^{\bar{a}\bar{b}\bar{c}}R_{\bar{a}de}^{a\bar{d}\bar{e}}
R_{\bar{b}\bar{d}f}^{bd\bar{f}}R_{\bar{c}\bar{e}\bar{f}}^{cef}
- R_{cef}^{\bar{c}\bar{e}\bar{f}}R_{bd\bar{f}}^{\bar{b}\bar{d}f}
R_{a\bar{d}\bar{e}}^{\bar{a}de}R_{\bar{a}\bar{b}\bar{c}}^{abc}
\end{equation}
\noindent to be zero. One such possibility corresponds to equation(13)
in Hietarinta's paper \cite{hieta2}, which is:

\begin{equation}
R_{111}^{222} = R_{222}^{111} = a, \\
R_{112}^{221} = R_{212}^{121} = b, \\
R_{121}^{212} = R_{212}^{121} = c, \\
R_{122}^{211} = R_{211}^{122} = d.
\end{equation}
\noindent where $a,b,c$ and $d$ are some arbitrary parameters with all
other entries of the R-matrix being zero.

\section{Technique}
\hspace*{0.2in} In this paper, most of the algebraic computations have
been done using the computer algebra, Mathematica \cite{wolf}.  The
method used is similar to that employed by Hietarinta in his analysis
of QYBE
\cite{hieta1}. Using a short program, we churn out all 256 equations
from FME, using suitably chosen ansatz. We then analyse the 256
equations for the unknowns.  These equations are generally  trivial,
though in the simplest case of a diagonal ansatz with one off-diagonal
element, there can be as many as 7  different ``quartic" equations with
nine unknowns.

The ansatz that we choose initially are basically diagonal ones or
cross-diagonal ones with increasing number of off-diagonal elements.
By systematically increasing the number of such off-diagonal elements,
we hope to push the list as far as possible.  We only manage to exhaust
all listing up till  two off-diagonal elements.  The task gets very
involved as the number of off-diagonal elements increases to three.  In
the case of two off-diagonal elements, there are still 1540 cases,
although this number can be substantially reduced by looking at the
discrete symmetries mentioned in section \ref{gensym}.  In
the case of three off-element, for instance, there are altogether 27720
possible cases which can of course be cut down easily by the symmetries
in the indices of the equations.

\section{Results}

\hspace*{0.2in} Solutions to FME are not always independent. Due to
invariance under the symmetry transformations, many solutions are
related to each other and the number of different solutions can largely
be reduced.

\subsection{Solutions with only one non-zero off-diagonal element}

\hspace*{0.2in} There are 56 possible off-diagonal positions for the
non-zero elements.  However, if we consider all possible discrete
symmetries, we need to consider only 12 different positions for the
non-zero off-diagonal element. A convenient choice of these positions
are shown in the array below:

\begin{equation}
\left( \begin{array}{cccccccc}
. & . & s_{1} & s_{2} & s_{3} & s_{4} & . & d_{1} \\
. & . & s_{5} & s_{6} & s_{7} & s_{8} &  d_{2} & . \\
. & . & . & . & . & d_{3} & . & . \\
. & . & . & . & d_{4} & . & . & . \\
. & . & . & . & . & . & . & .  \\
. & . & . & . & . & . & . & . \\
. & . & . & . & . & . & . & . \\
. & . & . & . & . & . & . & .
\end{array}
\right),
\end{equation}

In addition, it is found that the cases corresponding to the non-zero
element at the following positions $s_{7}$,
$s_{8}$ and $d_{3}$ respectively do not yield non-singular solutions.
Thus, we have effectively a total of nine different positions to
consider for the non-zero-element.

To present our solutions more compactly, we shall write the solutions
in the form

\begin{equation}
\{ R_{111}^{111}, R_{112}^{112}, R_{121}^{121}, R_{122}^{122},
R_{211}^{211}, R_{212}^{212}, R_{221}^{221}, R_{222}^{222} \},
\end{equation}

\noindent where  $R_{ijk}^{lmn}$ denotes the only non-zero off-diagonal
element. Further, without any loss of generality, we shall set
$R_{111}^{111}$ to unity.

The solutions are as follows:

\begin{enumerate}
\item $s_{1}: \hspace{3mm}$ $R_{111}^{121} = k \neq 0$
\medskip
\begin{equation}
\{ 1,  1 , 1 , a ,  1 ,  \frac{1}{a}, a , b \};
\end{equation}

There are also three other solutions in which complex entries occur:

\begin{equation}
\{ 1,  - \omega , - 1 , a ,  - \omega^{2} ,
\frac{\omega^{2}}{a}, - \omega^{2} a , b \};
\end{equation}

\noindent{where $\omega^3 = 1$. }

\item $s_{2}$: \hspace{3mm} $R_{111}^{122} = k \neq 0$

\medskip
\begin{equation}
\{ 1, \pm 1, \xi, \pm \xi^{6}, \pm \xi^{6} , \pm \xi, \xi^{12}, \pm
\xi^{2} \};
\end{equation}
\medskip
\begin{equation}
\{ 1, \zeta, \pm \zeta, -1, \zeta^{7}, \pm \zeta, -i, \zeta^{3} \};
\end{equation}

\noindent{where $\xi^{7} = \pm 1$ and $\zeta^8 = \pm 1$.}

\item $s_{3}$: \hspace{3mm}$ R_{111}^{211} = k \neq 0$
\medskip
\begin{equation}
\{ 1, a, b, c, -1, abd, d, e \};
\end{equation}

\noindent{where $b, d = \pm 1$.}

\item $s_{4}$: \hspace{3mm} $R_{111}^{212} = k \neq 0$
\medskip
\begin{equation}
\{ 1, a ,b ,\frac{1}{ab},a, \pm \frac{a^2}{b}, \frac{1}{ab}, \pm a^{2}b^{2} \},
\end{equation}

\begin{equation}
\{ 1, a ,\pm a^2 , \pm \frac{1}{a^3}, - a, -1 , \mp \frac{1}{b^3}, \mp b^6 \},
\end{equation}

\begin{equation}
\{ 1, a ,\pm a^2 , \pm \frac{1}{a^3}, ia, -1 , \mp \frac{i}{b^3}, \mp b^6 \},
\end{equation}

\begin{equation}
\{ 1, a ,\pm a^2 , \pm \frac{1}{a^3}, - ia, -1 , \mp \frac{i}{b^3}, \mp b^6 \}.
\end{equation}

\item $s_{5}$: \hspace{3mm} $R_{112}^{121} = k \neq 0$

This is a more complicated case than the rest and probably deserves
more discussion.  If we now write the solutions in the form:
\begin{equation}
\{ R_{111}^{111}, R_{112}^{112}, R_{121}^{121}, R_{122}^{122},
R_{211}^{211}, R_{212}^{212}, R_{221}^{221}, R_{222}^{222};
R_{112}^{121} \},
\end{equation}
the solution takes the form
\medskip
\begin{equation}
\{ 1,a,b,c,b,c,c,d;k \};
\end{equation}
where $a,b,c$ and $k$ are not independent, but related to each other
by the equations:

\begin{subeqnarray}
c^{2} &=& ab + ck; \\
(a^2 - 1) b +  k (a + b) & = & 0.
\end{subeqnarray}

Suppose we allow $b = c = 1$, we will get the solution:
\begin{equation}
\{ 1,a,1,1,1,1,d; 1 - a \}.
\end{equation}

Other possibilities exist.  If we allow $a = 1$ and $a = b$, we easily get:
\begin{equation}
\{1,1,-1,c,-1,c,c,d;  \frac{1 + c^2}{c} \};
\end{equation}
\noindent and
\begin{equation}
\{ c ~~ or ~ \frac{b^2}{c},b,b,c,b,c,c,d; \frac{c^2 - b^2}{c} \}
\end{equation}
\noindent respectively.

\item $s_{6}$: \hspace{3mm}$R_{112}^{122} = k \neq 0$

\begin{equation}
\{ 1, 1, 1, \omega ,\omega , a,1,\omega  \};
\end{equation}
\noindent where $\omega^{3} = 1$, $\omega \neq 1$.

\item $d_{1}$: \hspace{3mm} $R_{111}^{222} = k \neq 0$

There are 18 solutions but these solutions can be written compactly as:

\begin{equation}
\{ 1,  a, \lambda \omega^\prime a, \omega, a, \pm \omega^\prime, \omega,
\omega^\prime a   \};
\end{equation}
\noindent where $\omega^{3} = \omega^{\prime 3} = - 1$ and $\lambda^{2}
= \pm 1$.

\item $d_{2}$: \hspace{3mm} $R_{112}^{221} = k \neq 0$

\begin{equation}
\{ 1,a,\pm a, 1, a, \pm 1, 1, 1 \}.
\end{equation}

\item $d_{4}$: \hspace{3mm} $R_{122}^{211} = k \neq 0$
\begin{equation}
\{ 1,a,\pm a, 1, a, \pm 1, 1, 1 \}.
\end{equation}

\end{enumerate}
This completes the list of solutions possible under the diagonal ansatz
with one non-zero off-diagonal element.

\subsection{Solutions with two off-diagonal elements}

\hspace*{0.2in} There are $56 \choose 2$, i.e. 1540 possible positions.
Again, most solutions are related to each other by the similarity
transformation and discrete symmetries. In particular, there are two
interesting cases to consider:

\begin{description}
\item[Case 1:] The two non-zero elements are symmetrical about the
diagonal.
\item[Case 2:] The two elements are symmetrical about the cross diagonal.
\end{description}

\subsubsection{Case 1} \label{Case1}

\hspace*{0.2in} There are altogether four different types of
non-singular solutions.  They correspond, modulo the symmetry
transformations, to the cases when  $R_{111}^{112}$, $R_{111}^{121}$,
$R_{112}^{122}$, and $R_{112}^{212}$  and their respective entries by
reflection about the diagonal of the R matrix are non-zero, for
example, when $R_{111}^{112}$ and $R_{112}^{111}$ are non-zero
off-diagonal elements, and so forth.

If we denote the diagonal entries of the solutions by
\begin{equation}
\{ R_{111}^{111}, R_{112}^{112}, R_{121}^{121}, R_{122}^{122},
R_{211}^{211}, R_{212}^{212}, R_{221}^{221}, R_{222}^{222} \}
\end{equation}

\noindent we can present the solutions as:

\begin{enumerate}
\item Case when $R_{111}^{112}, R_{112}^{111}$ are not zero.
There are three distinct sets of soultions:
\begin{equation}
\{ 1,-1 ,a, \pm a , b, \pm  b, c, d  \} ,
\end{equation}
\noindent with $R_{111}^{112} = \dis{\frac{1 - a^2}{k}}$ and $R_{112}^{111}
= k$ ;
\begin{equation}
\{ 1, a, \pm (a + 1), \pm (a + 1), b, b, c, d \} ,
\end{equation}

\begin{equation}
\{ 1, a, \pm (a + 1), \mp (a + 1), b, -b, c, d \} ,
\end{equation}

\noindent with $R_{111}^{112} = \dis{\frac{a}{k}}$ and $R_{112}^{111} = k$.

\item Case when $R_{111}^{121}, R_{121}^{111}$ are not zero.
There are two sets of  solutions.  They  take the form:
\begin{equation}
\{ 1,  \pm 2, 1, a, \pm 2, \frac{4}{a},a,b \} ,
\end{equation}
\noindent{with $R_{111}^{121} = \frac{1}{k} $ or $\frac{-3}{k}$ and
$R_{121}^{111} = k $, and }

\begin{equation}
\{ 1,  a + 1, a, b, a + 1, \frac{(a + 1)^2}{b},b,c \} ,
\end{equation}
\noindent with  $R_{111}^{121} = \frac{a}{k} $ and $R_{121}^{111} = k $

\item Case when $R_{112}^{122}, R_{122}^{112}$ are not zero.

The solutions are:

\begin{equation}
\{ 1, a, b, 1 - a,1 , b, 1, c  \} ,
\end{equation}
\noindent {with $R_{112}^{122} = \dis{\frac{a(1 - a)}{k}}$ and $R_{122}^{112}
= k$, }

\noindent and

\begin{equation}
\{ 1, a, b, \frac{\pm \sqrt{b^2 - 4b} - b}{2},
\frac{- 4b + 2b^2 \pm 2b^{\frac{3}{2}} \sqrt{b - 4}}{4}, b, 1, c  \} ,
\end{equation}
\noindent {with $R_{112}^{122} = \dis{\frac{a(1 - a)}{k}}$ and $R_{122}^{112}
= k$.}

\item Case when $R_{112}^{212}, R_{212}^{112}$ are not zero.
There is one solution:
\begin{equation}
\{ 1, a, b, \frac{a + c}{b}, 1, c, b, {\dis \frac{a + c}{b} } \} ,
\end{equation}
\noindent with $R_{112}^{212} = \dis{\frac{ac}{k}}$ and $R_{212}^{112}
= k$

\end{enumerate}

These solutions seem to possess some regular patterns.
If we label the vertices of a cube by \{111\},\{112\},\{121\}, and so forth
as shown in figure 1, we make an interesting observation.  Suppose we
consider $R_{ijk}^{lmn}$ as the non-zero element and look at the
vertices on the cube corresponding to the indices $\{ijk\}$ and
$\{lmn\}$, we see that a non-singular solution exists only for cases in
which the vertices are connected by an edge of the cube.

\begin{figure}
\begin{picture}(500,200)
\multiput(140,0)(120,0){2}{\line(0,1){120}}
\multiput(140,0)(0,120){2}{\line(1,0){120}}
\multiput(140,0)(0,120){2}{\line(-1,1){60}}
\put(80,180){\line(1,0){120}}\put(80,180){\line(0,-1){120}}
\put(200,180){\line(1,-1){60}}
\multiput(200,180)(0,-20){6}{\line(0,-1){12}}
\multiput(80,60)(20,0){6}{\line(1,0){12}}
\multiput(255,5)(-30,30){2}{\line(-1,1){20}}
\multiput(140,0)(120,0){2}{\circle*{15}}
\multiput(140,120)(120,0){2}{\circle*{15}}
\multiput(80,60)(120,0){2}{\circle*{15}}
\multiput(80,180)(120,0){2}{\circle*{15}}
\put(150,10){$211$}\put(270,10){$221$}\put(150,100){$111$}
\put(270,100){$121$}\put(212,186){$122$}\put(92,190){$112$}
\put(92,65){$212$}\put(212,65){$222$}
\end{picture}
\caption{Labelling of Cube}
\end{figure}
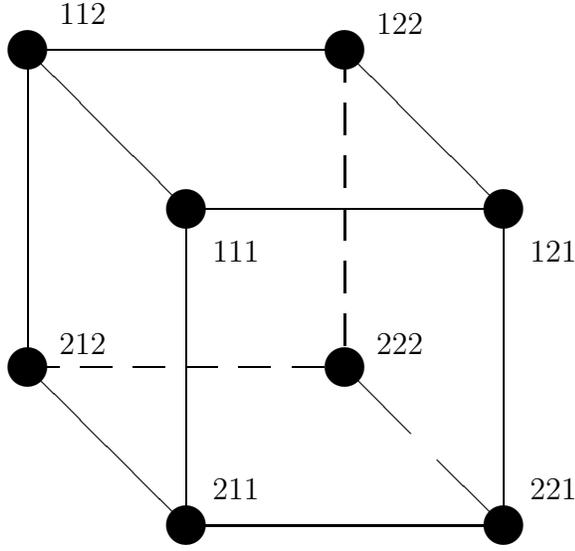
\vspace{0.5cm}

\subsubsection{Case 2}

\hspace*{0.2in} Non-singular solutions exist for the cases when
$R_{111}^{122}$, $R_{111}^{212}$, $R_{112}^{121}$, $R_{112}^{211}$,
$R_{111}^{112}$, $R_{121}^{122}$ and  $R_{112}^{212}$  and their
respective entries by reflection in the cross diagonal of the R matrix
are not zero. These solutions seem less interesting than the previous
cases.  The off-diagonal elements are in general independent of the
elements along the diagonal, except for case of solution (\ref{spec2}).

Up to symmetries, there are seven basic solutions.  Using the notations
in the previous subsection \ref{Case1}, the solutions are:
\begin{enumerate}
\item
\begin{equation}
\{ a, b, a, b, a, b, a, b \} ,
\end{equation}
\noindent with $R_{111}^{122} = c$ and $R_{211}^{222}
= d$ and $a^2 = b^2$;
\item
\begin{equation}
\{ a, b, a, b, b, a, b, a \} ,
\end{equation}
\noindent with $R_{111}^{212} = c$ and $R_{121}^{222}
= d$ and $a^2 = b^2$;
\item
\begin{equation}
\{ a, b, b, -b, b, -b, -b, a \} , \label{spec2}
\end{equation}
\noindent with $R_{112}^{121} = R_{212}^{221} = \dis{\frac{a^2 - b^2}{2a}} $;
\item
\begin{equation}
\{ a, a, a, -a, -a, a, a, a  \} ,
\end{equation}
\noindent with $R_{112}^{211} = b$ and $R_{122}^{221}
= c$;
\item
\begin{equation}
\{ a, -a, a, a, b, b, -b, b  \} ,
\end{equation}
\noindent with $R_{111}^{112} = c$ and $R_{221}^{222}
= d$;
\item
\begin{equation}
\{ a, a, b, b, b, b, a, a  \} ,
\end{equation}
\noindent with $R_{121}^{122} = R_{211}^{212}
= c$;
\item
\begin{equation}
\{ a, b, b, a, a, c, c, a  \} ,
\end{equation}
\noindent with $R_{112}^{212} = R_{121}^{221}
= d$;
\end{enumerate}
\subsubsection{Other Cases}

\hspace*{0.2in} Besides the cases mentioned, there are many solutions
which are not related to the above solutions by any symmetry
transformations mentioned in section \ref{gensym}. For example, in the
case in which $R_{111}^{112}$ is not zero and one other element
systematically chosen from the other 27 possible positions in the upper
triangle of the R-matrix is set as the non-zero element.  In this case,
we find solutions for cases in which

\begin{enumerate}
\item $R_{111}^{221} \neq 0 $: 2 possible solutions.
\item $R_{111}^{222} \neq 0$ : 2 possible solutions.
\item $R_{121}^{122} \neq 0$ : 4 possible solutions.
\item $R_{211}^{212} \neq 0$ : 4 possible solutions.
\item $R_{221}^{222} \neq 0$ : 8 possible solutions.
\end{enumerate}

A typical solution from this list appears as:

\begin{equation}
R =
\left( \begin{array}{cccccccc}
a & {\dis\frac{ac}{b} } & 0 & 0 & 0 & 0 & 0 & 0 \\
0 & -a & 0 & 0 & 0 & 0 & 0 & 0 \\
0 & 0 & \lambda a & 0 & 0 & 0 & 0 & 0 \\
0 & 0 & 0 & \mu a & 0 & 0 & 0 & 0 \\
0 & 0 & 0 & 0 & b & c & 0 & 0  \\
0 & 0 & 0 & 0 & 0 & -b & 0 & 0 \\
0 & 0 & 0 & 0 & 0 & 0 & \lambda b & 0 \\
0 & 0 & 0 & 0 & 0 & 0 & 0 & \mu b
\end{array}
\right),
\end{equation}

\noindent where $\lambda, ~ ~ \mu = \pm 1$.

\subsection{Cross diagonal ansatz with one or more non-zero elements }

\hspace*{0.2in} In comparison with the number of solutions which we can
generate by looking at diagonal R-matrices with one or more off
diagonal elements, there are fewer  solutions obtained from R-matrices
with non-zero cross diagonal elements and one or more elements off the
cross-diagonal. By cross diagonal ansatz, we refer to an R-matrix in
which  the elements
\begin{equation}
\{ R_{111}^{222}, R_{112}^{221}, R_{121}^{212}, R_{122}^{211},
R_{211}^{122}, R_{212}^{121}, R_{221}^{112}, R_{222}^{111} \}
\end{equation}
are not zero. We have shown earlier (see subsection \ref{cross})
that such R-matrix will satisfy FME when the parameters take on any
value.

It is interesting to note that such ansatz with one non-zero
off cross-diagonal element does not yield any non-singular solution.
Solutions exist only if the number of non-zero off cross-diagonal
elements exceeds unity. We shall describe one such class of solutions:
cross-diagonal ansatz with two non-zero elements placed symmetrically
about the cross-diagonal.

\subsubsection{Cross diagonal ansatz with two non zero elements}

\hspace*{0.2in} We shall list the cross diagonal elements as:

$$
\{ R_{111}^{222}, R_{112}^{221}, R_{121}^{212}, R_{122}^{211},
R_{211}^{122}, R_{212}^{121}, R_{221}^{112}, R_{222}^{111} \}
$$

For such an ansatz, we find that solutions exist essentially for two
cases:

\begin{enumerate}
\item Case 1: \hspace{10mm} $R_{111}^{122} = R_{211}^{222} = k$
The cross-diagonal elements are:
\begin{equation}
\{ a , b, \mu b, \lambda a, c, \frac{bc}{a}, \mu \frac{bc}{a}, \lambda
c \}
\end{equation}

\item Case 2:\hspace{10mm}  $R_{111}^{221} = R_{112}^{222} = k$ \\
The cross-diagonal elements are:
\begin{equation}
\{ \lambda a , \lambda b, \mu c, \mu \frac{bc}{a}, c,  \frac{bc}{a}, a
, b \}
\end{equation}
\end{enumerate}

\subsection{Other solutions}

\hspace*{0.2in} As we increase the number of off-diagonal elements, we
find increasing complexity in solving the 256 non-linear ``quartic"
equations.  The number of over-determined consistency equations
increases rapidly than the increase in the number of unknowns. A
typical result in which the four off-diagonal elements are symmetrical
about both the diagonals is:

\begin{equation}
R =
\left( \begin{array}{cccccccc}
a & a & 0 & 0 & 0 & 0 & 0 & 0 \\
b & b & 0 & 0 & 0 & 0 & 0 & 0 \\
0 & 0 & x & 0 & 0 & 0 & 0 & 0 \\
0 & 0 & 0 & x & 0 & 0 & 0 & 0 \\
0 & 0 & 0 & 0 & y & 0 & 0 & 0  \\
0 & 0 & 0 & 0 & 0 & y & 0 & 0 \\
0 & 0 & 0 & 0 & 0 & 0 & c & c \\
0 & 0 & 0 & 0 & 0 & 0 & d & d
\end{array}
\right),
\end{equation}
\noindent where $x^2 = (a + b)^2$ and $y^2 = (c + d)^2$.

One can easily check that Hietarinta's constant
upper triangular solution of ZTE \cite{hieta2},
\begin{equation}
\left(
\begin{array}{cc}
1 & k \\
0 & 1 \\
\end{array}
\right ) \otimes \left(
\begin{array}{cccc}
1 & p & q & r \\
0 & 1 & 0 & q \\
0 & 0 & 1 & p \\
0 & 0 & 0 & 1 \\
\end{array}
\right )
\end{equation}
satisfy FME. We have also found a similar upper triangular solution
which is not related to the above solution:

\begin{equation}
R =
\left( \begin{array}{cccccccc}
1 & 0 & 0 & 0 & p & 0 & 0 & q \\
0 & 1 & 0 & 0 & 0 & p & -p & 0 \\
0 & 0 & 1 & 0 & 0 & -p & p & 0 \\
0 & 0 & 0 & 1 & p^{2} q^{-1} & 0 & 0 & p \\
0 & 0 & 0 & 0 & 1 & 0 & 0 & 0  \\
0 & 0 & 0 & 0 & 0 & 1 & 0 & 0 \\
0 & 0 & 0 & 0 & 0 & 0 & 1 & 0 \\
0 & 0 & 0 & 0 & 0 & 0 & 0 & 1
\end{array}
\right),
\end{equation}

There is also an interesting ``bi-diagonal" solution analogous to
Zamolodchikov's ``bi-diagonal" spectral dependent solution.  It appears
as:

\begin{equation}
R =
\left( \begin{array}{cccccccc}
a & 0 & 0 & 0 & 0 & 0 & 0 & p \\
0 & a & 0 & 0 & 0 & 0 & q & 0 \\
0 & 0 & \pm b & 0 & 0 & q & 0 & 0 \\
0 & 0 & 0 & a & r  & 0 & 0 & p \\
0 & 0 & 0 & q & b & 0 & 0 & 0  \\
0 & 0 & r & 0 & 0 & \pm a & 0 & 0 \\
0 & r & 0 & 0 & 0 & 0 & a & 0 \\
qrp^{-1} & 0 & 0 & 0 & 0 & 0 & 0 & b
\end{array}
\right),
\end{equation}

\section{Conclusion}

\hspace*{0.2in} As noted in Frenkel and Moore's original paper, the FME
may be a good strategy to investigate Zamolodchikov's tetrahedron
equation. The tetrahedron equation has so far admitted only one
well-known solution, which is the original spectral dependent solution
proposed by Zamolodchikov himself. On the other hand, there is a wealth
of solutions, albeit spectral free, which we can generate from Frenkel
and Moore's equation. By systematically increasing the number of
unknowns in the R-matrix, we may be able to discover some symmetries
which are inherent in both the tetrahedron equation and FME. Recently,
Hu
\cite{hu}  has attempted to relate the FME to braid groups. More
recently, Li and Hu has shown that a given representation of the braid
group induces a special kind of solution for the FME \cite{hu2}. They
have also invoked symmetry transformations (\ref{sym2} b) and
(\ref{sym3} c) in their solution. It would seem therefore that a
systematic understanding of FME will help us gain greater insights into
higher dimensional generalization of QYBE.

\end{document}